\documentclass[a4paper,10pt]{iopart}

\usepackage{graphicx}

\begin{document}

\title[Suppression of Mott--Hubbard states and
metal-insulator transitions ...]{Suppression of Mott--Hubbard states and
metal-insulator transitions in the two band Hubbard model}

\author{J.P.Hague\dag\ddag}
\address{\dag\, Dept. of Physics and Astronomy, University of Leicester, Leicester, LE1 7RH, U.K. }
\address{\ddag\, Max-Planck Institut f\"{u}r Physik Komplexer Systeme, Dresden 01187, Germany}

\begin{abstract}
I investigate band and Mott insulating states in a two-band Hubbard
model, with the aim of understanding the differences between the
idealised one-orbital model and the more realistic multi-band
case. Using a projection ansatz I show that additional orbitals
suppress the metal-insulator transition, leading to a critical
coupling of approximately eight times the bare band-width. I also
demonstrate the effects of orbital ordering, which hinder Mott-Hubbard
states and open a band gap. Since multi-band correlations are common
in real materials, this work suggests that the existence of very
strongly correlated band insulators may be more common than
Mott-Hubbard insulators. [PUBLISHED AS J.PHYS.:CONDENS. MATTER {\bf 17} 1385-1397 (2005)]
\end{abstract}


\section{Introduction}
\label{section:introduction}

The prediction and experimental confirmation of a correlation driven
Mott transition is one of the great success stories from the study of
correlated electron systems. Mott postulated the existence of an
insulating state, when the Coulomb repulsion is significantly larger
than the kinetic energy, as the antithesis of the free-electron metal
\cite{mott}. The simplest model of the Mott transition is the Hubbard
model, where a one-band tight-binding model is supplemented with a
site-local repulsion between electrons of opposite spin, which
partially represents the Coulomb interaction \cite{hubbard1963}. The
one-band Hubbard model is insulating at half-filling, provided that
the interaction between electrons is significantly higher than the
inter-site hopping, since electrons may not move freely without paying
a large energy penalty.

The single-band Hubbard model is very effective for systems in which
the electrons at the Fermi-energy are well separated from those in
other bands. In many systems, including the much discussed cuprate
superconductors, it is clear that two or even more bands are close in
energy and inter-band coupling is expected to be relevant. In
particular, electrons may favour configurations associated with Hund's
rule coupling between orbitals. The aim of this paper is to understand
the differences between one- and two-band Hubbard models, including
the effects of realistic orbitals and orbital ordering. It is
particularly important to classify this behaviour, since
band-structure theories tend to play down the r\^{o}le of strong
correlations, while many-body theorists are inclined to over-simplify
their models.

Orbital ordering and weak-correlation Hartree-Fock effects are the
typical origins of band-gaps in electronic structure calculations. The
name \emph{band insulator} is given, because in the event of weak
interactions, the gap separates \emph{single-electron bands}. An
alternative mechanism for the opening of band-gaps is found for very
strong inter-electron coupling, where electron correlations drive an
insulating state at half-filling (exactly two electrons per site in
this paper). The origin of the strongly-interacting \emph{Mott-Hubbard
insulator} is far more complicated in origin, and the gap is formed by
many-body interactions between electrons of the same type and may be
found in systems with strong correlation. A full classification of
different insulating states may be found e.g. in reference
\cite{gebhard}.

The treatment of excitations in the two-band Hubbard model presented
here involves a local ansatz-based approach to the correlation
problem. For the ground state, an approximation which treats local
spin and density fluctuations is used. Such an approach has been
applied to $d$-electrons in transition metals by Stollhoff \emph{et
al.}  \cite{stollhoff}. Excited states are also treated using a
local spin and density fluctuation approximation using the approach of
Becker \emph{et al.}  \cite{becker}. The present work modifies the
method of Unger \emph{et al.} \cite{unger}, although in this paper we
are interested in the physics of the two-band Hubbard model for a
range of interactions, rather than the specific details of real
metallic systems like Nickel.

This paper is organised as follows. The model is presented in section
\ref{sec:model} along with a discussion of the non-interacting
problem, and details of the metal to band insulator transition. In
section \ref{sec:excited}, the effects of correlation on the orbitally
degenerate model are calculated for a range of Coulomb and Hund's
coupling. In section \ref{sec:bandmott}, the effects of changing the
orbital degeneracy on correlated states are discussed. Conclusions are
presented in section \ref{sec:conclusions}. In particular, two-band
effects have a dramatic influence on the Mott transition and change
the critical coupling by an order of magnitude.

\section{Model and formalism}
\label{sec:model}

\begin{figure}
\begin{indented}\item[]
\includegraphics[width=50mm]{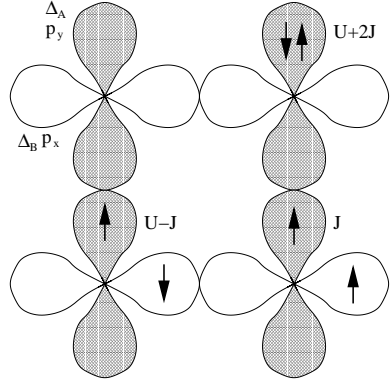}
\end{indented}
\caption{In the model used in this work, the first set of $p$-orbitals
extend in the $x$-direction, while the second set extend in the
$y$-direction. In addition to the orbitals, the energy levels of the
orbitals represented by $\Delta_i$ may be changed, adding a competing
mechanism for an insulating state to the system. Electrons interact
via a Coulomb repulsion, and inter-orbital spins interact via a Hund's
rule coupling, $J$. For two electrons of opposite spin in the same
orbital, the repulsion is $U+2J$, and for electrons in different
orbitals, it is $U-J$. }
\label{fig:model}
\end{figure}

\begin{figure}
\begin{indented}\item[]
\includegraphics[width=70mm,height=100mm,angle=270]{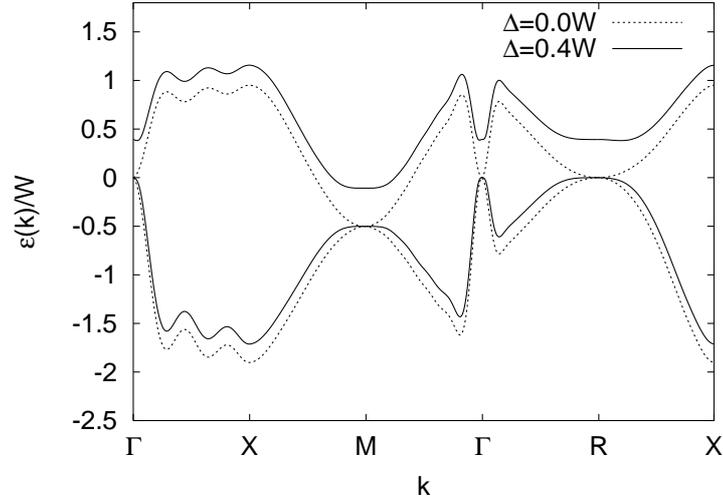}
\end{indented}
\caption{Non-interacting band structure resulting from diagonalisation
of the canonical structure matrix. The orbitals are strongly
hybridised and there are two clearly separated bands. There are
several degenerate regions along the high symmetry directions. Also
shown is the non-interacting band structure as the energy level of one
of the orbitals is raised by $0.4W$. This removes the degeneracies at
high-symmetry points. At a critical energy difference of
$\Delta=1.43W$, the bands split completely, and the system is driven
through a metal-to-band insulator transition. It should be stressed
that \emph{band insulators} have very different properties to
correlation driven \emph{Mott insulators}, which are the main subject
of this paper.}
\label{fig:barebandstructure}
\end{figure}

\begin{figure}
\begin{indented}\item[]
\includegraphics[width=70mm,height=100mm,angle=270]{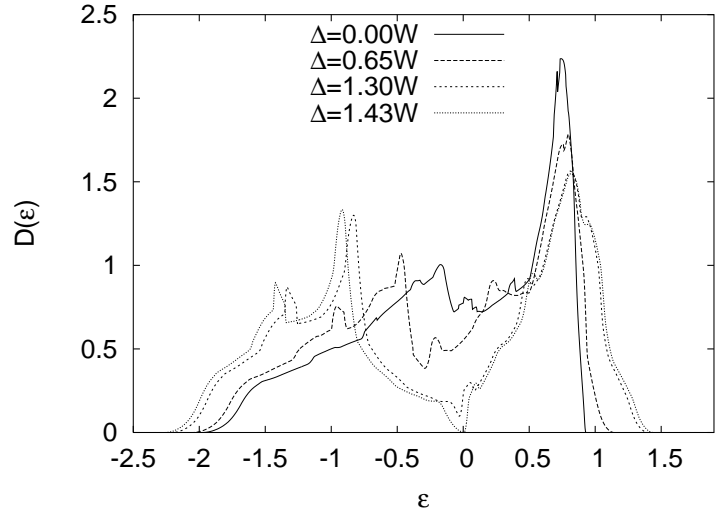}
\end{indented}
\caption{The bare density of states resulting from the canonical
structure matrix. There is no particle-hole symmetry. Also shown is
the effect of changing the relative energy levels of the
orbitals. Precursors to the band are visible just below the Fermi
energy. The opening of a band-insulating gap can be seen at
$\Delta=1.43W$.}
\label{fig:baredos}
\end{figure}

The model used in this work describes electrons on a simple cubic
lattice of interlocking $p$-orbitals, and is shown schematically in
figure \ref{fig:model}. Two bands are considered, with one orbital
aligned along the $x$-axis and the other along the $y$-axis. This
model has the property that hopping within the plane is equivalent
along both axes, with a different (smaller) hopping along the
$z$-axis. As such, the model may be considered to describe a quasi-2D
system. Electrons may be created in the $m$th band with the operator
$d^{\dag}_{mk\sigma}$. Each band has a Coulomb repulsion, $U$, and
spins from different bands interact with each other via a local
coupling, $J$. The most general form of model for treating locally
interacting electrons is,
\begin{eqnarray}
H=&\sum_{m\mathbf{k}\sigma}(\epsilon_{m\mathbf{k}}+\epsilon_F)d^{\dag}_{m\mathbf{k}\sigma}d_{m\mathbf{k}\sigma}\nonumber\\
&+(U+2J)\sum_{iI} n_{iI\uparrow}n_{iI\downarrow} + \frac{1}{2}(U-J)\sum_{I\sigma\sigma',i\neq j}n_{Ii\sigma}n_{Ij\sigma'}\\
&-J\sum_{I,i\neq j}\mathbf{s}_i.\mathbf{s}_j\nonumber
\end{eqnarray}

The non-interacting (or bare) band structure is found by diagonalising
the one-electron canonical structure matrix \cite{andersen},
\begin{equation}
S^{\mathbf{k}}_{l'm';lm}=g_{l'm';lm}\Sigma^{\mathbf{k}}_{\lambda,\mu}
\end{equation}
where
\begin{equation}
\Sigma^{\mathbf{k}}_{\lambda,\mu}=\sum_{R\neq 0}e^{i\mathbf{k}.\mathbf{R}}\left(\frac{S}{R}\right)^{\lambda+1}[\sqrt{4\pi}i^{\lambda}Y_{\lambda\mu}(\mathbf{\hat{R}})]^{*}
\end{equation}
and
\begin{eqnarray}
g_{l'm';lm}=2(-1)^{m+1}&\left(\frac{(2l'+1)(2l+1)}{(2\lambda+1)}\right)^{1/2}\\
&\times\left(\frac{(\lambda+\mu)!(\lambda-\mu)!}{(l'+m')!(l'-m')!(l+m)!(l-m)!}\right)^{1/2}\nonumber
\end{eqnarray}

$Y_{\lambda\mu}(\mathbf{R})$ are the spherical harmonics, and $\lambda=l'+l$,
$\mu=m'-m$. After diagonalisation, the Eigenvalues represent the
dispersion of the non-interacting bands, and Eigenvectors transform
the problem from the band to the orbital representation. The model
represented by the canonical-structure matrix is more complicated that
that typically used for Hubbard models, since hopping is allowed
between all sites rather then just nearest neighbours. The largest
contributions are expected from nearest and next-nearest neighbours,
although other smaller corrections are expected from longer range
hopping.

One may also augment the model by shifting the energies of individual
orbitals,
\begin{equation}
H_0\rightarrow H_0+\sum_{Ii\sigma}\Delta_i n_{Ii\sigma}
\end{equation}
resulting in an orbitally diagonal matrix that is simply added to the
canonical structure matrix,
\begin{equation}
\Delta_{\lambda,\mu}=\Delta_{\lambda}\delta_{\lambda,\mu}
\end{equation}

The resulting band structure for the two-band $p$-orbital model is
shown in figure \ref{fig:barebandstructure}. As with all figures, the
Fermi energy is at $\epsilon=0$. For a model with degenerate
orbital energies, the bands hybridise into upper and lower
bands. There are several degenerate points along the high symmetry
directions. Modifying the orbital energies has the effect of
separating the upper and lower bands, creating local band gaps. It can
be seen in the figure that the highest energies of the lowest band are
higher than the lowest of the higher band, and there is no universal
band gap. The band that is lower in energy is favoured by
electrons. As $\Delta$ is increased, the local band gaps become
larger, until there is no overlap and the band gap is universal across
the Brillouin zone. Then, a metal-to-band insulator transition takes
place at half filling, since the ground state favours one completely
filled and one empty band.

I turn to the evolution of the band insulating state in figure
\ref{fig:baredos}. The total DOS is calculated using the analytic
tetrahedron method \cite{lambin}. As $\Delta$ is increased, the
precursor of the band gap can be seen just below the Fermi energy at
$\epsilon=0$. The gap finally opens at $\Delta=1.43W$, where the band
width $W=\sqrt{M_2-M_1^2}$ is taken as the band width of the bare
($\Delta=0$) dispersion, and $M_n$ is the $n$th moment of the DOS.

Such a band insulator is trivially formed, and effects of
non-degenerate orbital energies are expected to be found in a wide
range of materials. However, it is the effects of correlation that are
the main subject of this paper. When one includes a simple Hubbard
interaction in the lower band, it is clear that some electrons will be
excluded from the lower orbital, and the band-insulating state will be
destroyed. From the inverse viewpoint, the effects of correlation
are hindered by a band state. In the remaining sections of this
paper, I will investigate first correlations, and then the effects of
band insulating states on these correlations.

\section{Excited states and the Mott transition}
\label{sec:excited}

In this section I examine whether there is a Mott transition in the
two $p$-orbital model, and what its nature is. The method for studying
correlated states used here is based on a projection ansatz, where the
effects of local spin and density fluctuations are considered. Further
details of this method can be found in reference \cite{unger}. In
brief, the current approach is to calculate the Green's function using
the projection method of Mori \cite{mori} and Zwanzig
\cite{zwanzig}. The propagator is projected onto the space spanned by
operators $\{B_1$...$B_N\}$, and the Green's function may then be
calculated as,
\begin{equation}
\underline{G}(\mathbf{k},\omega)=\underline{X}[\omega\underline{X}-\underline{F}]^{-1}\underline{X}
\label{eqn:greensfunction}
\end{equation}
with
\begin{equation}
F_{\mu\nu}=\langle\Omega|[B^{\dagger}_{\mu},{\mathcal{L}} B_{\nu}]_{+}\Omega\rangle_c
\end{equation}
\begin{equation}
X_{\mu\nu}=\langle\Omega|[B^{\dagger}_{\mu}, B_{\nu}]_{+}\Omega\rangle_c
\end{equation}
Note that $\underline{F}$ and $\underline{X}$ are
Hermitian. $\mathcal{L}$ is the Liouville operator, defined by
$\mathcal{L}O=[H,O]_{-}$. The state
$|\Omega\rangle_c=|1-\sum_{\mu}\eta_{\mu}A_{\mu}\rangle_c$ is a
correlated ground state where the effects of local two-particle
excitations have been projected out of the Hartree-Fock state. The
subscript $c$ indicates that these values should be constructed as
cumulants (i.e. $\langle A|B\rangle_c=\langle A^{\dag}B\rangle-\langle
A^{\dag}\rangle\langle B\rangle$). This avoids considering unnecessary
statistically independent processes. For full details about the
cumulant formalism, please refer to Reference \cite{fulde}. Once
matrix \ref{eqn:greensfunction} has been calculated, the element
$G_{00}(\mathbf{k})$ contains the electron Green's function. The poles
of this function define the quasi-particle dispersion.

Following Unger \cite{unger}, the $A$ operators are defined as,
\begin{eqnarray}
A^{1}_{ij}&=2\delta n_{iI\uparrow}\delta n_{iI\downarrow} & \,(i=j) \label{eqn:densfluc1}\\
&=\delta n_{iI}n_{jI} & \,(i> j) \label{eqn:densfluc2}\\
A^{2}_{ij}&=\mathbf{s}_{iI}.\mathbf{s}_{jI} & \,(i> j)\label{eqn:spinfluc}
\end{eqnarray}
in order to take into account local spin (equation \ref{eqn:spinfluc}) and density (equations \ref{eqn:densfluc1} and \ref{eqn:densfluc2}) fluctuations. $B$
operators create an additional electron accompanied by a fluctuation
(spin or density) and are defined as,
\begin{eqnarray}
B^{0}&=d^{\dagger}_{iI\uparrow} & \\
B^{1}_{ij}&=2d^{\dagger}_{iI\uparrow}\delta n_{iI\downarrow} &\,(i=j)\\
&=d^{\dagger}_{iI\uparrow}\delta n_{jI}&\,(i\neq j)\\
B^{2}_{ij}&=\frac{1}{2}(d^{\dagger}_{iI\uparrow}S^{z}_{jI}+d^{\dagger}_{iI\downarrow}S^{+}_{jI})&\,(i\neq j)\\
B^{3}_{ij}&=\frac{1}{2}d^{\dagger}_{jI\downarrow}d^{\dagger}_{jI\uparrow}d_{iI\downarrow}&\,(i\neq j)
\end{eqnarray}
In this way, a total of nine $B$ operators are defined for the
two-band model. Full details of the calculation of the matrix elements
required to compute the Green's function in equation
\ref{eqn:greensfunction} are found in reference \cite{unger}.

\begin{figure}
\begin{indented}\item[]
\includegraphics[width=50mm,height=100mm,angle=270]{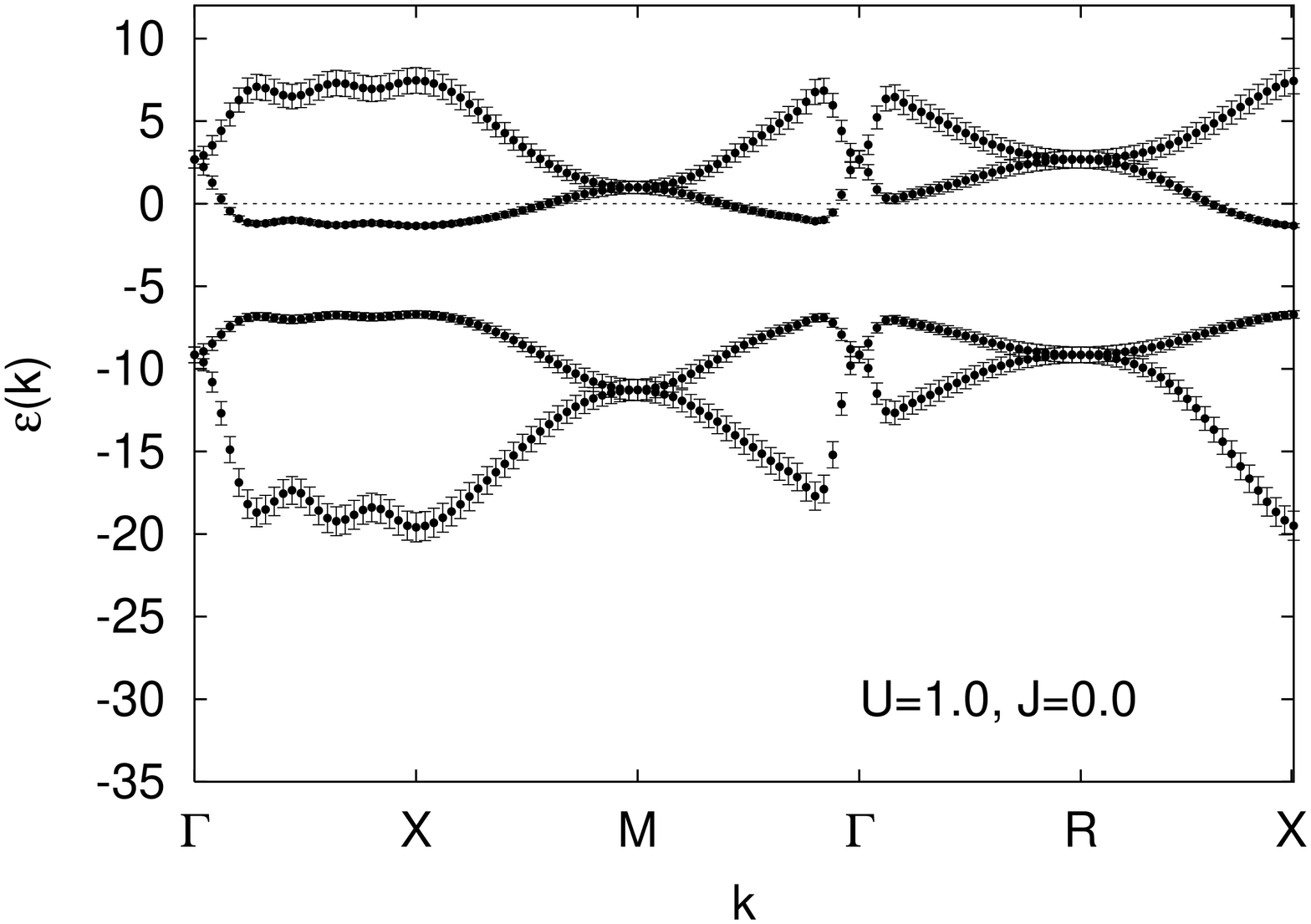}
\includegraphics[width=50mm,height=100mm,angle=270]{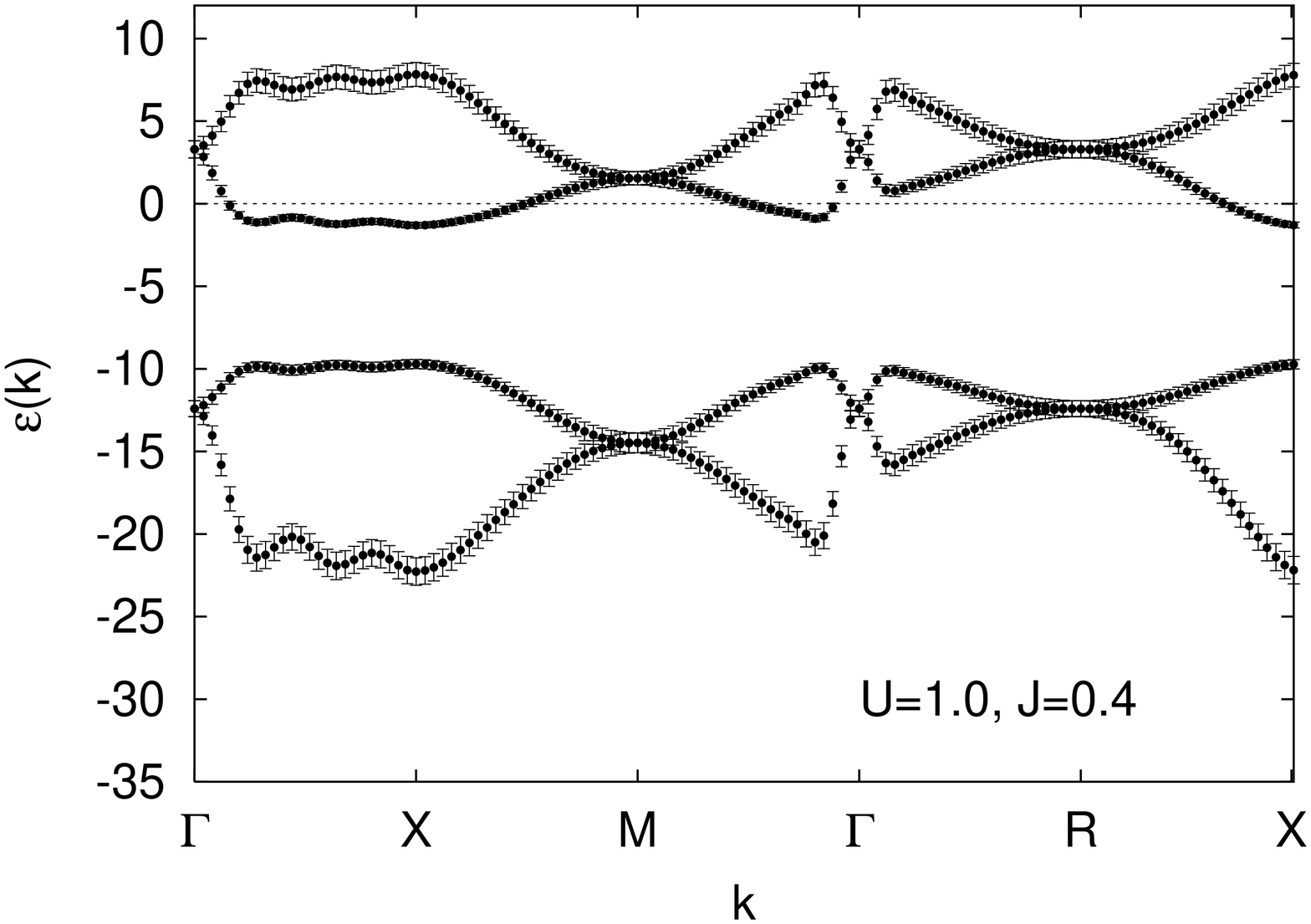}
\includegraphics[width=50mm,height=100mm,angle=270]{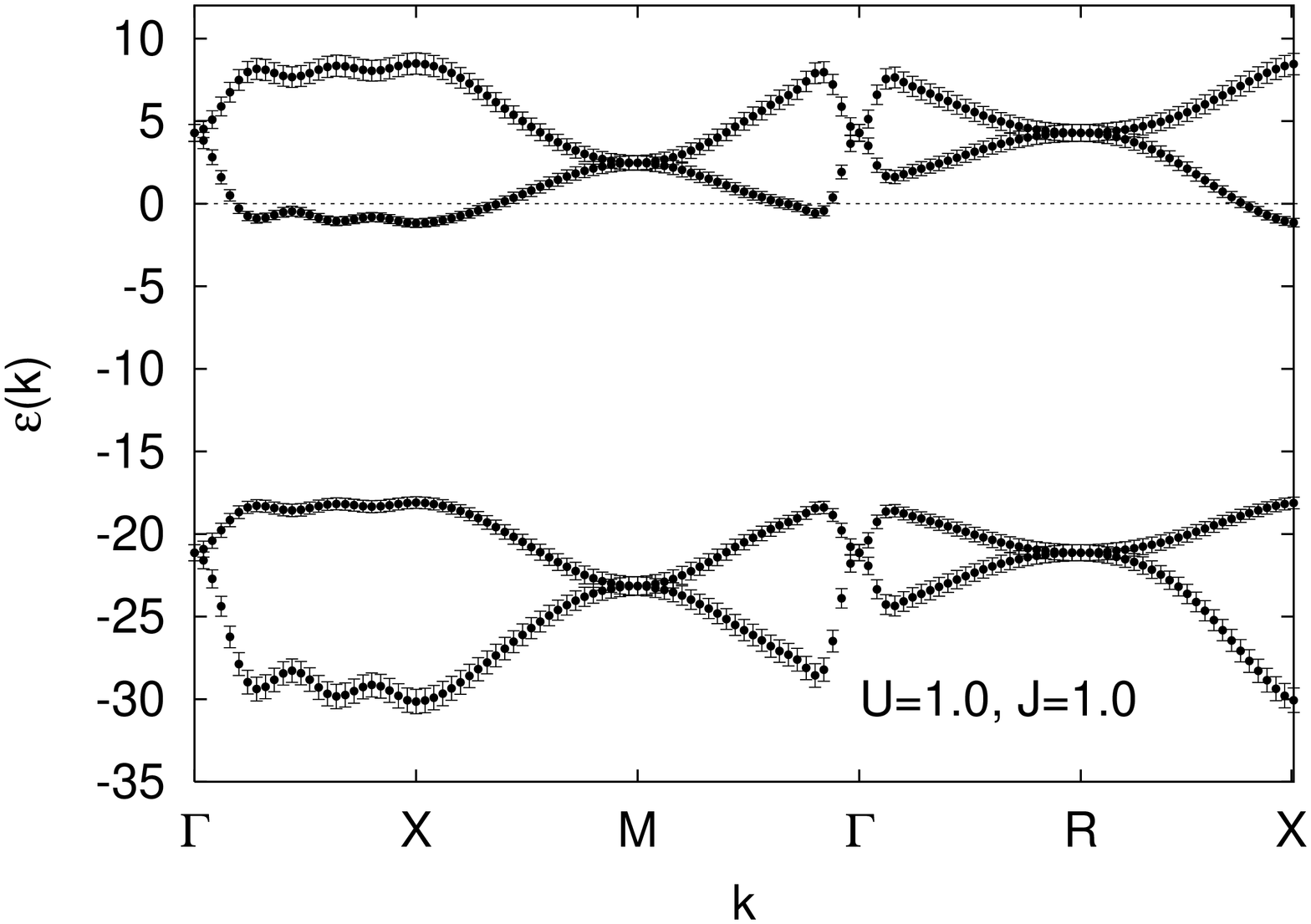}
\end{indented}
\caption{Effect of coupling on the bare band structure: (a) U=W, J=0,
(b) U=W, J=0.4W and (c) U=W, J=W. The momentum path is taken along the
high symmetry directions. ``Errorbars'' on the plot indicate the
residue or quasi-particle weight of the band. The Fermi energy is
indicated by the light dotted line. Note that unlike in the simple one
band case that corresponds to the Hubbard III approximation
\cite{hubbard}, bands do not split into sub-bands with equal
weight. The only constraint is that $Z_1+Z_3=1$ and $Z_2+Z_4=1$. This
is important, since the lower sub-bands are made up of two different
weights, and the metal-insulator transition need no longer exist at
half-filling. The metal insulator transition will occur when $Z_1=Z_4$
and (as implied) $Z_2=Z_3$. The effect of increasing $U$ and $J$ is
that $Z_1\rightarrow Z_4$. The lower and upper sub-bands narrow, and
the Mott gap grows larger. At the transition, the insulating gap is
discontinuous.}
\label{fig:effectofu}
\end{figure}

In figure \ref{fig:effectofu}, I show the effects of correlations on
the band structure of the two band Hubbard model when $U=W$ for
various $J$. The correlated dispersion is shown along the main
symmetry directions. In can be seen that two almost identical copies
of the original dispersion are formed (the top sub-band are
approximately the mirror image of the bottom sub-bands in the $x$-axis
of the graph). The differences between the one and two band models are
immediately clear. In the one band case, these bands would split into
two identical parts with equal weight, and for sufficiently high $U$
the bands would untangle and there would be a correlation driven metal
insulator (or Mott) transition. In the two-band case, it is no longer
necessary that each band splits into lower and upper sub-bands of
equal weight. One can see that the weights of the lower sub-band and
the second highest sub-band are related in the sense that $Z_1+Z_3=1$
and similarly $Z_2+Z_4=1$, where the subscripts indicate the order of
the band from low to high energies. Both geometric effects and more
importantly inter-band correlation effects are clearly at work here,
since a double-occupied (DO) triplet state can be created in the two
band model, allowing the lowest band to absorb the majority of the
electrons and for the total energy to be significantly lowered (such a
state is prohibited in the one-band model due to the Pauli exclusion
principle). The weights $Z_1+Z_2\rightarrow Z_3+Z_4$ as coupling
increases. From bottom to top, the sub-bands are related to: (1) DO
electron states with spins in parallel directions, (2) DO electron
states with spins in opposite directions, (3) antiparallel DO holes
and (4) parallel DO holes. Taking this into account, it is clearly
much more difficult to obtain an insulating state in the two-band
model, since the weights of the splitting must become symmetric before
this takes place.

\begin{figure}
\begin{indented}\item[]
\includegraphics[width=70mm,height=100mm,angle=270]{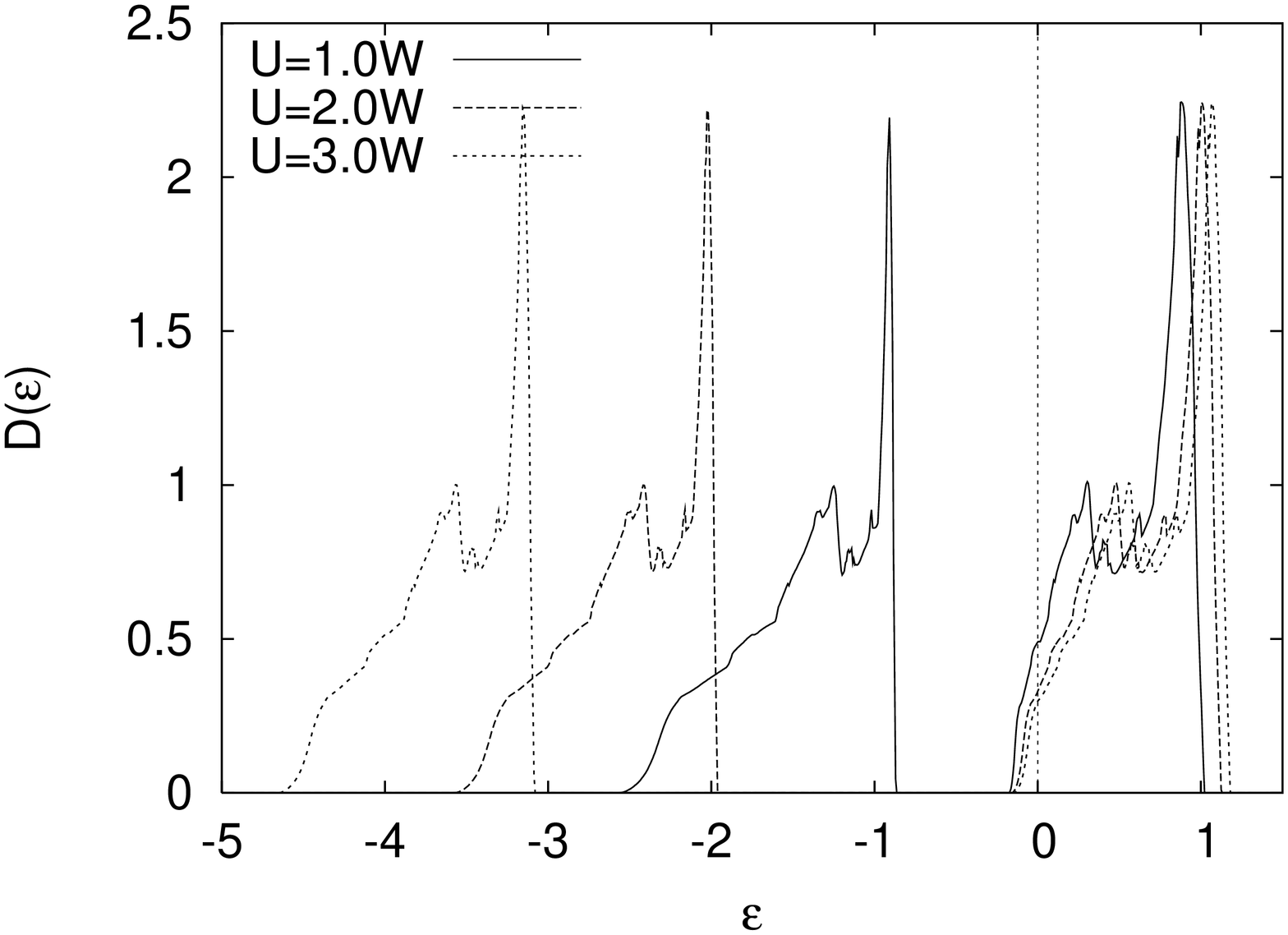}
\end{indented}
\caption{Effect of coupling on the DOS. $J$ is set to zero. The
increase in $U$ effects a widening of the band gap, and of the bands,
although the band width, $W_{\mathrm{upper}}$ calculated from the
moments decreases on average. The Mott-transition occurs at
$U=8W$. Note that an extremely small $U$ can open a gap in the density
of states. However, the weight of lower and upper bands is not equal,
so the low $U$ gap does not lead to an insulating state. The origin of
the low $U$ gap is almost certainly due to the nature of the model,
which to lowest order may be considered to represent interlocking
chains, and therefore has some 1D characteristics.}
\label{fig:dosvaruj0}
\end{figure}

\begin{figure}
\begin{indented}\item[]
\includegraphics[width=70mm,height=100mm,angle=270]{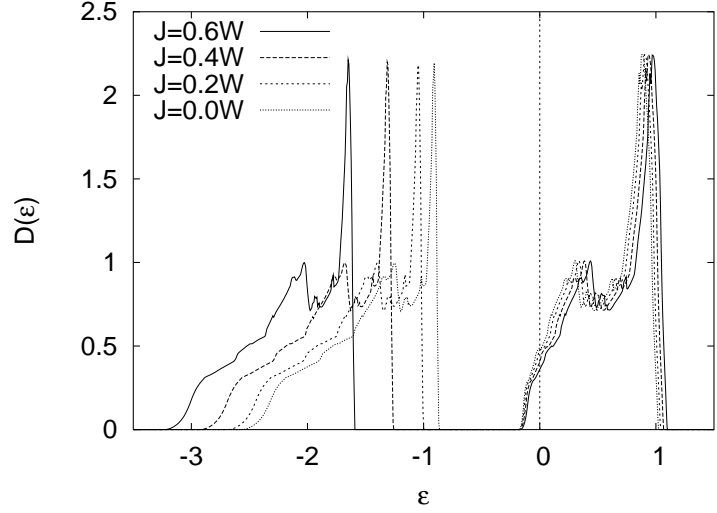}
\end{indented}
\caption{Effect of exchange on the DOS. $U=W$ and $J$ is varied. The
energy is in units of the bandwidth. As $J$ increases, the Mott gap is
increased. However, the lower two sub-bands contain less than two
electrons, and the Fermi energy lies within the top two sub-bands. As
such there are states at the Fermi-surface. A Mott transition is
expected at higher (unphysical) $J$.}
\label{fig:dosvarj}
\end{figure}

Figure \ref{fig:dosvaruj0} shows the effect of coupling on the DOS
when $J$ is set to zero. The increase in $U$ causes a widening of the
gap, which is not positioned at the Fermi surface ($\epsilon=0$). The
ratio of $W_{\mathrm{upper}}$ to $W_{\mathrm{lower}}$ steadily tends
to one. At $U=8W$, the system goes through a dramatic phase
transition. Owing to the slightly 1D nature of the $p$-orbital chains,
a band gap appears for very small $U<W$. This is not the
Mott-insulating gap (which opens later), but certainly has its origin
in correlation effects. The opening of the Mott gap will be discussed
below. Similar effects are seen for positive $J$. In figure
\ref{fig:dosvarj}, the effect of spin-spin coupling on the excited
states is shown for $U=W$. As $J$ increases, the Mott gap is
increased. However, the lower two sub-bands contain less than two
electrons, and the Fermi energy lies within the top two sub-bands. As
such there are states at the Fermi-surface, and for low $J$ the system
is metallic.

\begin{figure}
\begin{indented}\item[]
\includegraphics[width=70mm,height=100mm,angle=270]{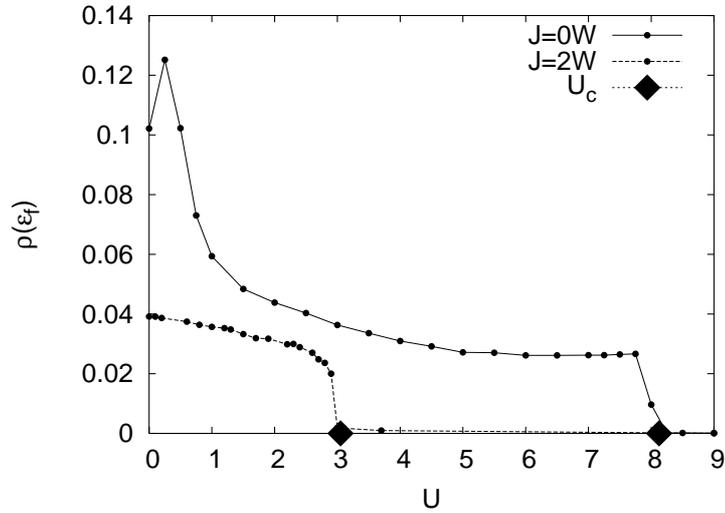}
\end{indented}
\caption{Density of states at the Fermi surface $\rho(\epsilon_f)$ vs
$U$ at half filling with $J=0$ and $J=2W$. For the simple model with
$J=0$, the onset of the Mott transition is greatly suppressed, with
the transition occurring at $U\sim8W$, more that double that expected
for the one-band model. Introduction of the spin-spin coupling $J$
suppresses the DOS at the Fermi surface, pushing the system closer to
the Mott state. For $J=2W$, a Mott transition can be seen at
$U=3W$. The very small $\rho(\epsilon_f)$ above $U=3W$ is due to
small numerical errors.}
\label{fig:rhoef}
\end{figure}

In order to investigate the nature of the Mott transition, I plot the
density of states at the Fermi surface, $\rho(\epsilon_F)$ in figure
\ref{fig:rhoef}. The upper curve shows the evolution of this property
for $J=0$. With slightly increasing $U$, $\rho(\epsilon_F)$ increases
slightly. This is due to the geometry of the bands, and non strictly
due to correlation. Above $U=0.25W$, the DOS at the Fermi surface
begins to fall, tending to a constant value at around $U=6W$, where
presumably the weights of the upper and lower sub-bands have reached a
constant value. This persists up to $U=8W$. Then the weights of the
upper and lower bands undergo a sudden change so that
$W_{\mathrm{upper}}=W_{\mathrm{lower}}$ and a transition from metal to
Mott-Hubbard insulator takes place. For $J=2W$, the transition occurs
at a much lower value of $U=3W$, and the weights of the upper and
lower sub-bands constantly change up to the transition, since the
strong coupling limit has not yet been reached. The critical
couplings, $U_C$ are shown as large diamonds.

\begin{figure}
\begin{indented}\item[]
\includegraphics[width=70mm,height=100mm,angle=270]{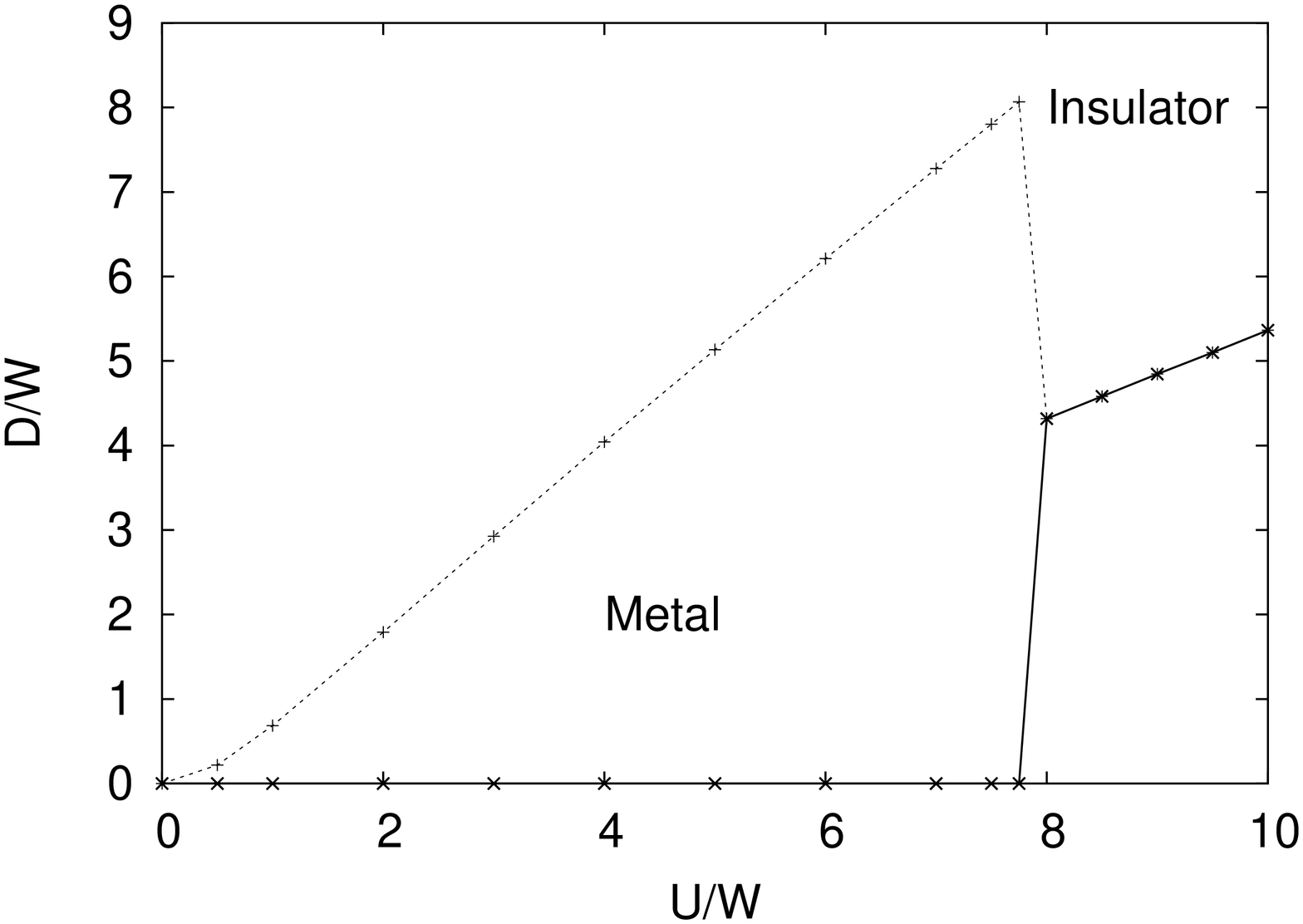}
\end{indented}
\caption{Mott gap when $J=0$. In the metallic state, a gap (D) forms
just below the Fermi energy. At $U_c$, a gap shifts to the
Fermi-surface. At the same time, the Mott gap halves. The opening of
the Mott gap at the Fermi surface is discontinuous at $U=8W$.}
\label{fig:gap}
\end{figure}

As an additional demonstration of the unusual nature of the
Mott-Hubbard transition in this model, I plot the magnitude of the gap
in the DOS at $J=0$ in figure \ref{fig:gap}. Starting from very small
$U$, a gap in the DOS opens approximately linearly in $U$ up to $U_c$
(as shown in the faint dotted line). The low $U$ gap is not positioned
at the Fermi-surface, so the system is metallic, although not expected
to be a good conductor. As previously stated, it is probable that a
low $U$ gap opens since in the non-interacting model, the system of
$p$-orbitals acts in a similar way to crossed 1D chains. At the
critical coupling $U_c$, however, the size of the gap suddenly reduces
and moves to the Fermi-surface. The system goes through a metal
insulator transition. The gap at the Fermi-surface is shown as a thick
solid line. The transition is unusual, since the opening of the gap is
discontinuous.

\begin{figure}
\begin{indented}\item[]
\includegraphics[width=70mm,height=100mm,angle=270]{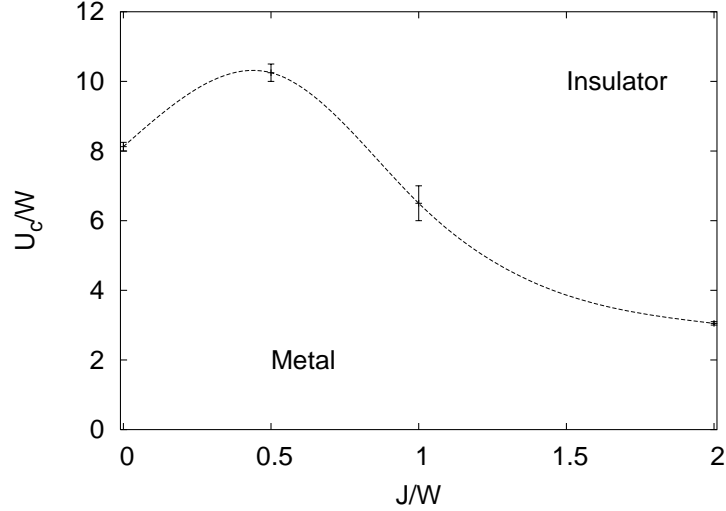}
\end{indented}
\caption{Phase diagram of the two band model at half filling. The
lines are a guide to the eye. It can be seen that the critical
coupling $U_c$ is significantly increased in the low $J$ two-band
case. It is expected that this trend will continue as more bands are
introduced.}
\label{fig:phasediagram}
\end{figure}

In summary, and to close this section, I have investigated the
emergence of Mott-Hubbard states in a two-band Hubbard model with
degenerate $p$-orbitals. The Mott transition is found at a significant
$U_c=8W$ for the $J=0$ model where the inter- and intra- band
couplings are equal. This is in agreement with the DMFT results
published in references \cite{kawakami,koga} for non-orbital based models where
the hopping for different bands may be changed. In reference
\cite{kawakami}, this increase is attributed to enhanced orbital
fluctuations. In this work, such fluctuations can be seen as an
asymmetric assignment of weights to certain double occupied and vacant
states. As $J$ is increased, the DOS at the Fermi surface is
suppressed. For $J=2W$ there is a correlation driven Mott
metal-insulator transition at $U_c=3W$. Both of these transitions have
the unusual characteristic that the insulating gap opens
discontinuously. These results are summarised in figure
\ref{fig:phasediagram}.

\section{Orbital anisotropy and band-Mott effects}
\label{sec:bandmott}

By changing the relative orbital energies, I also investigate the
band-Mott transition. Such a transition is particularly important in
the study of correlated electron systems, since one should understand
the competition between trivial band insulating effects such as those
found in semiconductors, and more complex interaction-driven Mott
insulators.

\begin{figure}
\begin{indented}\item[]
\includegraphics[width=35mm,height=100mm,angle=270]{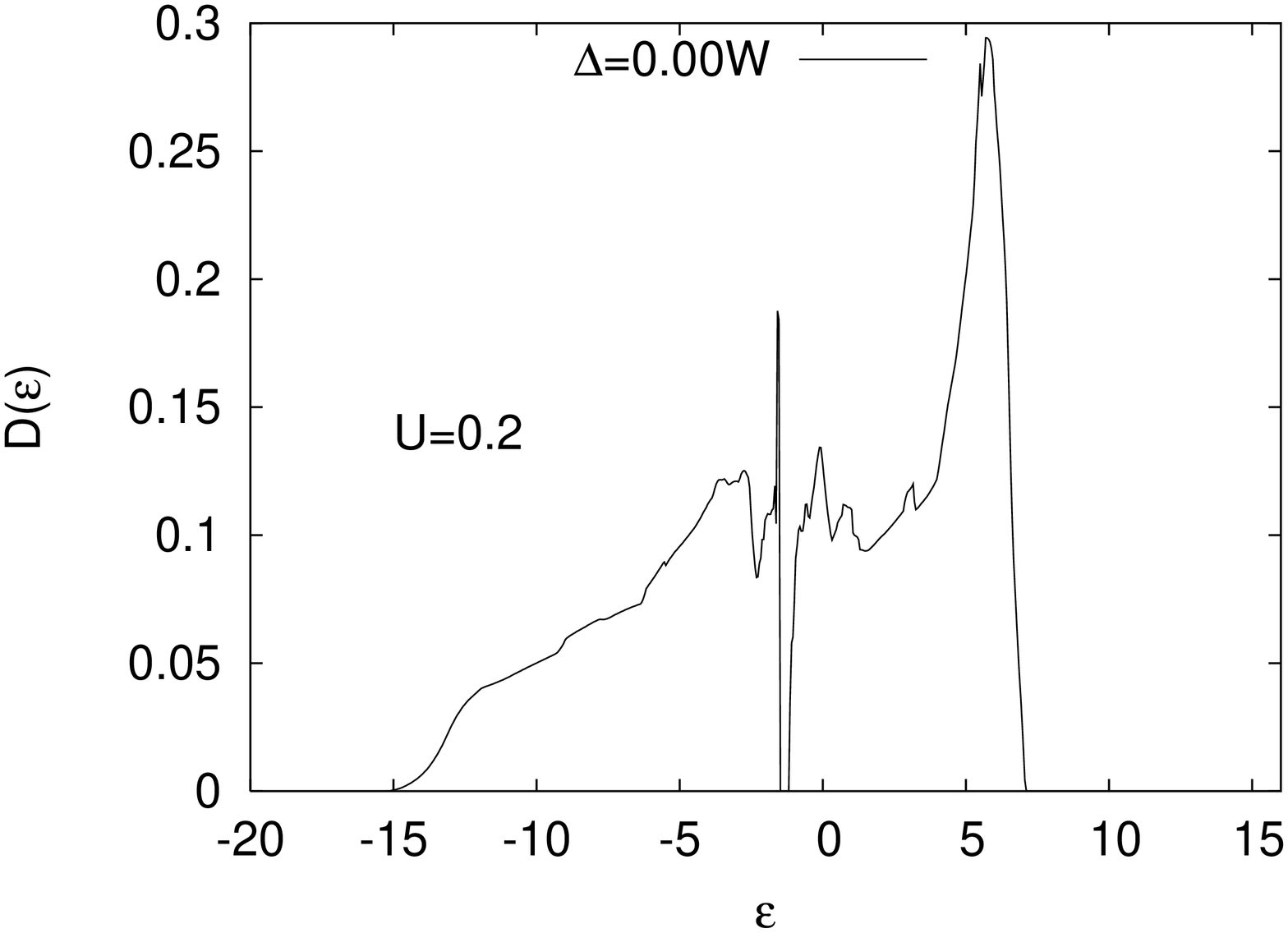}
\includegraphics[width=35mm,height=100mm,angle=270]{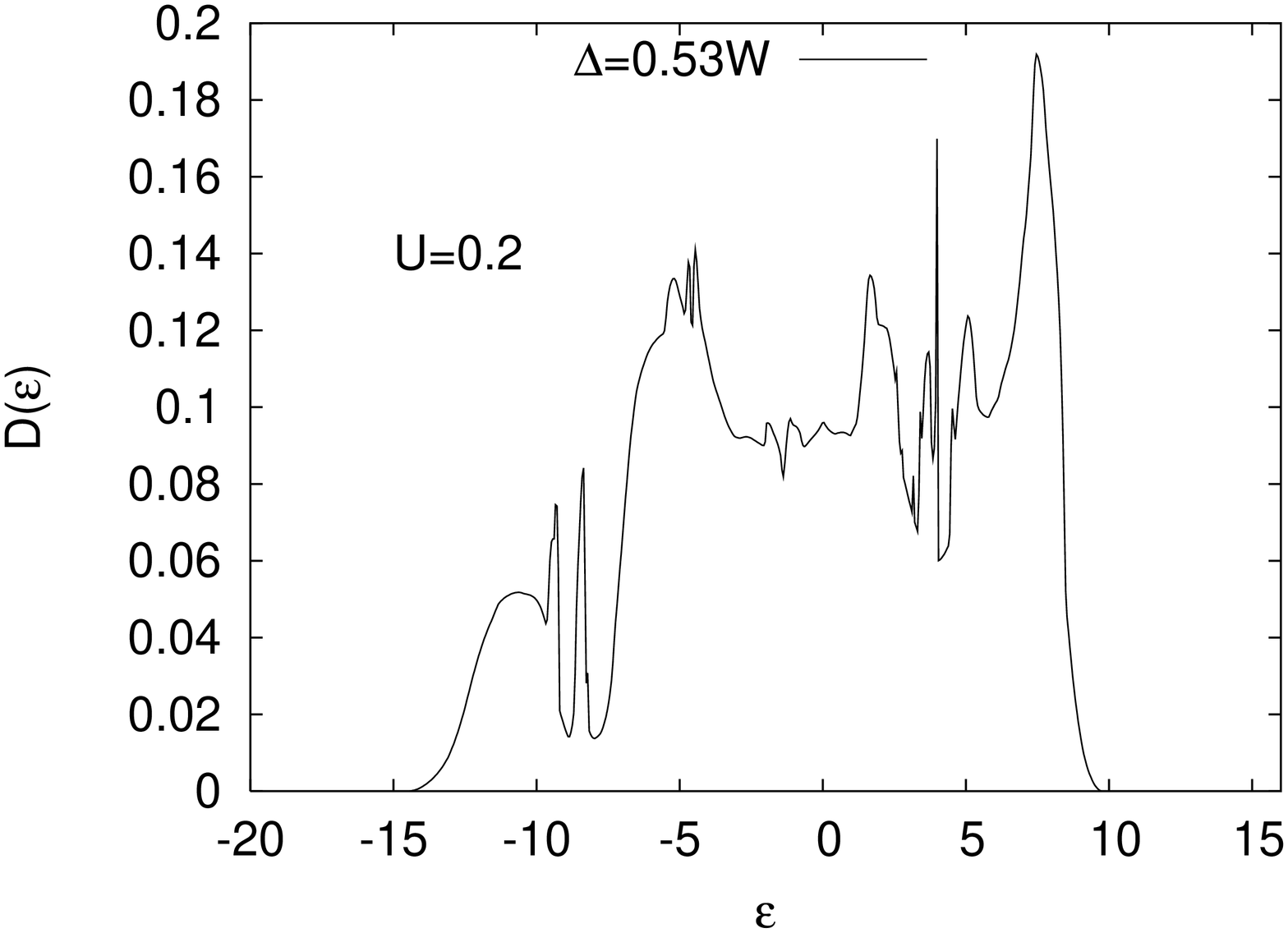}
\includegraphics[width=35mm,height=100mm,angle=270]{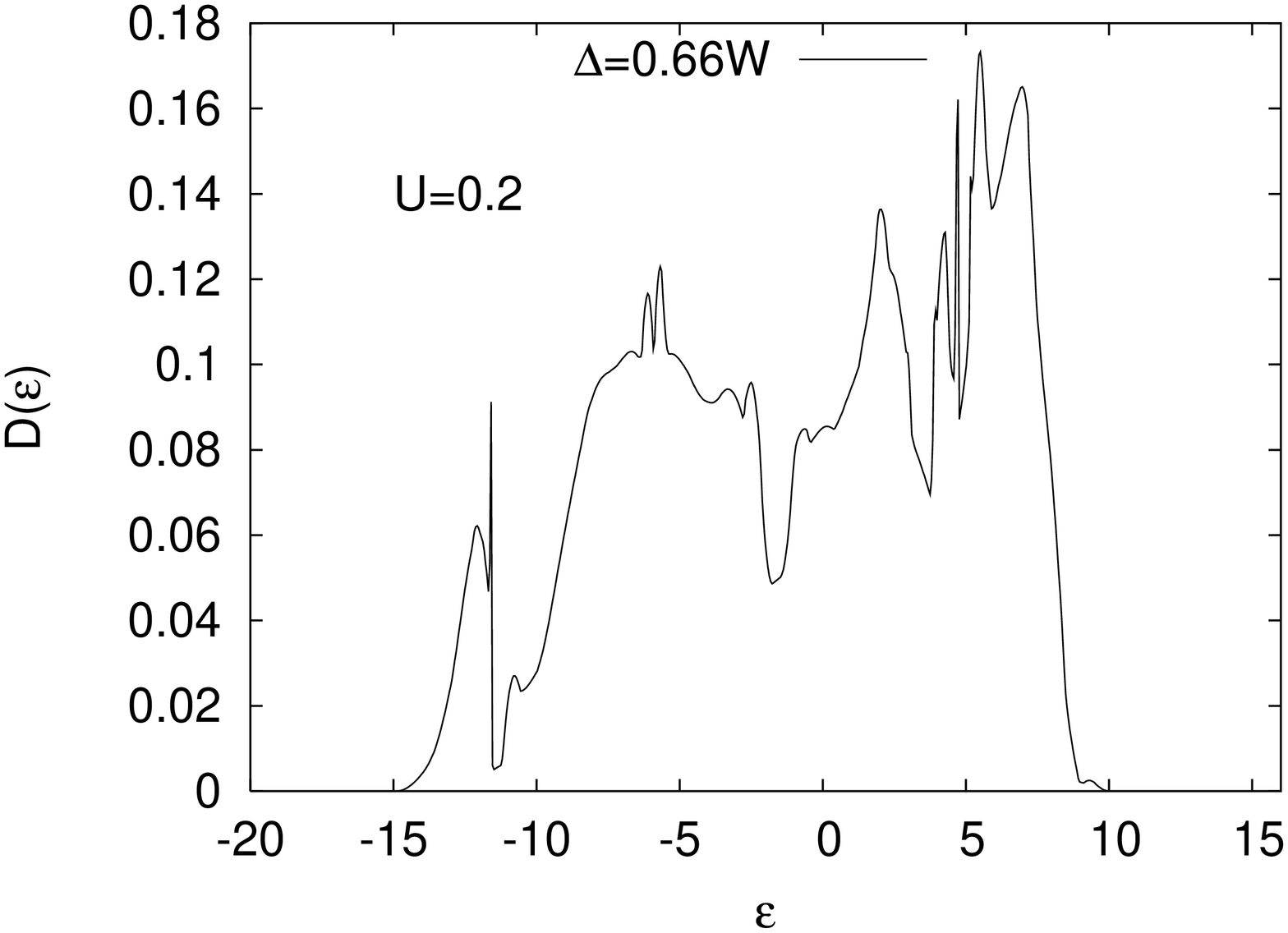}
\includegraphics[width=35mm,height=100mm,angle=270]{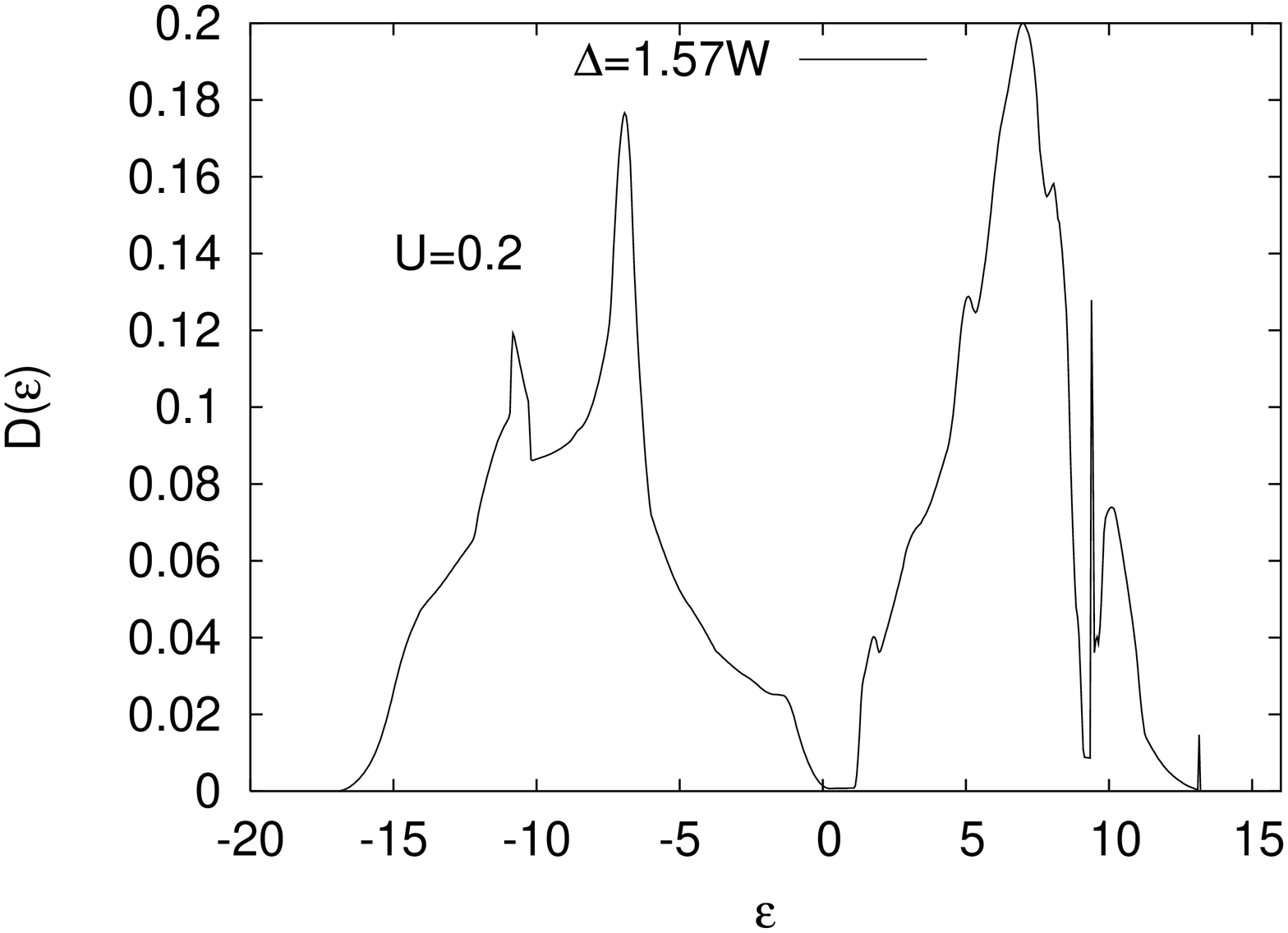}
\end{indented}
\caption{Effect of varying orbital energies on the DOS of the 2 band
system. (1) Degenerate orbital energies, (2) $\Delta=0.53W$, (3)
$\Delta=0.66W$ and (4) $\Delta=1.57W$. Each system is calculated with
$U=0.2W$. In the top panel a gap (which is not at half-filling)
forms. Varying orbital energies acts against a Mott gap. For
$\Delta=0.53W$, one can see that there is no gapped state. For
$\Delta=0.66W$, orbital ordering starts to dominate, and a gap begins
to form close to the Fermi-surface. This gap continues to grow, until
the system passes through a metal- band insulator transition between
$\Delta=1.5W$. The opening of the gap is
continuous in this case.}
\label{fig:bandmotttransition}
\end{figure}

Figure \ref{fig:bandmotttransition} shows the effect of varying
orbital energies on the DOS of the 2 band system when $U=0.2W$. The
top panel shows degenerate orbital energies, and then moving down, the
bands are separated by $\Delta=0.53W$, $\Delta=0.66W$ and
$\Delta=1.57W$.

In the top panel a Mott gap (which is not at half-filling)
forms. Varying orbital energies acts against this Mott gap. For
$\Delta=0.53W$, the conflicting energy scales are similar, and one can
see that there is no gapped state. The reason for the disappearance of
the gapped state is clear. The lowering of the energy of a single
orbital makes the orbital favorable for electrons, so that double
occupied states will form. The introduction of a Coulomb repulsion to
that orbital means that one of those electrons will be pushed out of
that orbital into the higher orbital, and as such electrons can hop
again, and a metal results. For $\Delta=0.66W$, orbital ordering
starts to dominate, and a gap begins to form close to the
Fermi-surface. This gap continues to grow, until the system passes
through a metal to band insulator transition at $\Delta=1.5W$.

The metal to band insulator transition is more easily understood,
since all the states associated with a band are pushed up in energy
until there is a phase transition. Since the states associated with a
single band must conserve particle number even in the presence of
correlation, then a weight, $W_{1}=1$ is pushed up, while a weight
$W_{2}=1$ is pushed down. It is therefore clear that for sufficiently
large $\Delta$, a metal-insulator transition is guaranteed at
half-filling. From this mechanism, it is also clear that the gap will
always open continuously with respect to $\Delta$.

\section{Conclusions}
\label{sec:conclusions}

I have studied the two-dimensional Hubbard model using a projection
ansatz that treats the effects of local spin and density
fluctuations. A simple model involving two p-orbitals aligned along
the $x$- and $y$-axes was solved, with an on-site Coulomb repulsion
$U$ combined with inter-orbital spin-spin coupling $J$. In addition to
electron-electron interactions, the relative orbital energies were
changed by a quantity $\Delta$ to open a band gap. Since paramagnetic
states are considered, correlation-driven insulating states are
Mott-Hubbard like \cite{gebhard}.

I found that the critical coupling of the Mott transition is much
larger in the two-band model than for a simple one-band system. In
particular, $U_{c}=8W$ for the two-band case with $J=0$, which is much larger
than the simple condition for the one-band model that $U$ should be
greater than the band-width to untangle the bands and open a
gap. These results are largely in agreement with the work of Kawakami
\cite{kawakami}. The coupling $J$ acts to suppress $U_{c}$. For all
$J$, the Mott gap opens discontinuously. I also examined the effects
of orbital ordering on the Mott states. The orbital ordering energy
$\Delta$ acts first to close the Mott gap, leading first to an
uncorrelated metal, and finally opening a band-insulating gap at
$\Delta\sim 1.43W+U$.

A non-insulating gap in the DOS is found for low $U$. In this regime,
the conduction electrons (i.e. those at the Fermi-surface) are found
to be associated with only one of the bands. In this sense, the small
$U$ transition could be identified with a primary Mott transition
associated with band A which is fustrated by conducting electrons in
band B. Such a transition is known as an orbitally selective Mott
transition \cite{anisimov2002a}. I note that the sub-band gaps at
small $U$, are of approximately the same size for both bands, which
doesn't completely rule out the possibility that the transition is
concurrent \cite{liebsch2003a}. However, reference \cite{liebsch2003a}
used a model of non-interacting electrons that neglected hybridisation
effects. The inclusion of hybridisation effects makes the model
significantly more complicated, and probably leads to the difference
in results.

Since a large Coulomb repulsion is required to form an insulating
state, even in the case of significant inter-orbital spin-spin
coupling, generating Mott-Hubbard states is clearly a difficult task
when dealing with real systems beyond the simplified physics of one
electron band. The physical origin of this difficulty is that while in
the ideal one band model, the band splits into two identical parts
with weight $W_{\mathrm{lower}}=W_{\mathrm{upper}}=\frac{1}{2}$, this
is no longer assured when geometrical effects from real orbitals
(including hybridisation), and correlation effects from multiple bands
are taken into account. The only constraint is that the total weight
of particles in each set of sub-bands (e.g. all sub-bands resulting
from band A) is conserved,
i.e. $W_{A,\mathrm{lower}}+W_{A,\mathrm{upper}}=1$. It is expected
that these effects will become more pronounced as the number of bands
increases. Inducing a band insulating state introduces none of these
problems, since a whole band (with total
$W_{\mathrm{lower}}+W_{\mathrm{upper}}=1$) is shifted down in energy,
until bands no longer overlap, and an insulator is assured.

How do these results relate to real systems? In most materials,
especially oxides, several electron bands are important, and orbital
energies are shifted due to a combination of geometric effects from
the crystal formation and the Hartree--Fock effects of interaction
with distant bands (which are well separated in energy). It is
therefore important that the effects of correlations in multi-band
models competing with orbital energy shifts are well understood. On
the basis of the evidence presented in this paper, it is apparent that
it is difficult to open a Mott insulating gap in a two-band material,
since the critical coupling is so high for realistic values of the
spin-spin coupling $J/U\sim0.1$, yet on the other hand, opening band
gaps is relatively easy. It is indeed possible that the existence of
very strongly correlated band insulators is much more common. Clearly
more work should be carried out to assess this situation.

For example, a realistic model of transition metal oxides would take
into account the $e_g$ orbitals which describe the transient electrons
in systems such as the CMR manganites. These have the form
$\cos(\theta/2)|x^2-y^2\rangle\pm
i\sin(\theta/2)|3z^2-r^2\rangle$. Also, further lowering of symmetries
can change the inter-site hopping along different axes, to generate
quasi 1D and 2D systems. Finally, the effect of increasing the number
of bands should be considered. Such calculations will form the focus
of future work.

%
%

\ack

I am pleased to acknowledge hospitality and support from the MPIPKS
guest scientist program and the Condensed Matter Physics Group at the
University of Leicester. I would especially like to thank Prof. Peter
Fulde for drawing my attention to this problem, and for extremely useful
discussions.

\vskip 10mm

\bibliographystyle{unsrt}
\bibliography{twobandhubbard}

\end{document}